\RequirePackage{fix-cm}
\documentclass[twocolumn,epjc3]{svjour3}  
\smartqed  
\RequirePackage{graphicx}
\RequirePackage{latexsym}
\RequirePackage[numbers,sort&compress]{natbib}
\RequirePackage[colorlinks,citecolor=blue,urlcolor=blue,linkcolor=blue]{hyperref}
\journalname{Eur. Phys. J. E}
\usepackage{amssymb}
\usepackage{amsmath}
\usepackage{tikz}
\usetikzlibrary{arrows.meta}
\usepackage{bbm}
\usepackage{dsfont}

\begin{document}

\title{A one-dimensional three-state run-and-tumble model with a `cell cycle'
}


\author{Davide Breoni\thanksref{e1,addr1}
        \and
        Fabian Jan Schwarzendahl\thanksref{addr1} 
        \and
        Ralf Blossey\thanksref{addr2} 
        \and
        Hartmut L\"owen\thanksref{addr1}
}

\thankstext{e1}{e-mail: breoni@hhu.de}


\institute{Institut f\"ur Theoretische Physik II: Weiche Materie, Heinrich-Heine-Universit\"at D\"usseldorf, Universit\"atsstra{\ss}e 1, 
40225 D\"usseldorf, Germany \label{addr1}
           \and
           University of Lille, Unit\'e de Glycobiologie Structurale et Fonctionnelle (UGSF), CNRS UMR8576, 59000 Lille, France \label{addr2}
}

\date{Received: date / Accepted: date}

\abstractdc{
{We study a one-dimensional three-state run-and-tumble model motivated by the bacterium {\it Caulobacter crescentus} which displays a cell cycle
between two non-proliferating mobile phases and a proliferating sedentary phase. Our model implements kinetic transitions between the two mobile
and one sedentary states described in terms of their number densities, where mobility is allowed with different running speeds in forward and backward direction.
We start by analyzing the stationary states of the system and compute the mean and squared-displacements for the distribution of all cells, as well as
for the number density of settled cells. The latter displays a surprising super-ballistic scaling $\sim t^3$ at early times.
Including repulsive and attractive interactions between the mobile cell populations and the settled cells, we explore the stability of the system and 
employ numerical methods to study structure formation in the fully nonlinear system. We find traveling waves of bacteria, whose occurrence is quantified in a 
non-equilibrium state diagram. 
}}

\maketitle

\section{Introduction}

Understanding the motion of bacteria has been a classic problem of biophysics \cite{berg_e_2004,berg_random_2018}. Bacteria are propelled by their flagellae, whose 
motor generates a torque which translates into forward or backward motion of the bacteria. The problem has also found interest within the soft matter community, as 
bacteria are but one example of a much larger class of systems, commonly denoted as microswimmers \cite{elgeti_physics_2015}. 
The run-and-tumble (RT) model of an active particle system is originally motivated by specific features of bacterial motion: this motion 
only persists for a finite time, the `run'-time, after which the bacterium stalls, the `tumble'-period, before continuing its motion typically in a different direction, 
see e.g. \cite{polin_chlamydomonas_2009}.
The properties of the basic RT model have been confronted with experiments, e.g. in \cite{detcheverry_generalized_2017,fier_langevin_2018}. 
The RT model also relates to other stochastic processes, e.g. the exclusion process \cite{bertrand_dynamics_2018} or even to the dynamics of quantum particles 
\cite{maes_diffraction_2022}. 

RT models in one dimension are a special case within this model class. Here, the bacterium can only switch between left- and right motion in a stochastic
manner. One-dimensional RT-models have proven to be an extremely rich field for analytic calculations; exemplary papers dealing with diverse aspects are:
confinement \cite{angelani_confined_2017}; space-dependent velocities, space-dependent transition rates and general drift velocity distributions 
\cite{angelani_run-and-tumble_2019,dhar_run-and-tumble_2019,singh_local_2021,monthus_large_2021,frydel_generalized_2021}; hard-core particles with spin 
\cite{dandekar_hard_2020}; inhomogeneous media \cite{singh_run-and-tumble_2020}; attractive/repulsive interactions 
\cite{le_doussal_stationary_2021,barriusogutierrez_collective_2021}; phase transitions \cite{mori_first-order_2021}; entropy production \cite{frydel_intuitive_2022}. 
Field-theoretic methods have been applied to RT models recently as well \cite{garcia-millan_run-and-tumble_2021,zhang_field_2022}.

In some sense, the (one-dimensional) RT model can be thought of playing in active systems a role analogous to Ising models in equilibrium statistical mechanics.
In the very recent past, several works have appeared carrying this analogy further, since they consider the number of `states' in which the bacterium can find
itself to go beyond the dichotomy of left- and right-moving states. Models with three and even more states have been discussed - in our Ising-model analogy, 
this amounts to looking at active analogues of `Potts'-type models \cite{basu_exact_2020,grange_run-and-tumble_2021,frydel_run-and-tumble_2022}. 

The present paper inserts itself in this line of research by considering a three-state RT model with the states: left-moving, right-moving and sedentary. Our model
is motivated by the behavior of the bacterium {\it Caulobacter crescentus} ({\it CC}), a model organism in microbiology since it has a complex lifestyle 
\cite{shebelut_caulobacter_2010,  biondi_cell_2022}. {\it CC} has a bacterial analogue of a cell cycle usually found in eukaryotes; in order to undergo cell division, 
the bacterium has to switch from its mobile swarmer state to a spatially localized stalked state. Only from the latter state the proliferation of new cells is possible. 
Our model, capturing this biological feature, is however not limited to {\it CC} or bacteria alone. 
E.g., the green algae \textit{Chlamydomonas reinhartii} has a similar cell cycle \cite{harris_chlamydomonas_2009} with 
sedentary and swimming states and also performs a run-and-tumble motion \cite{polin_chlamydomonas_2009}. 
The capacity of cell division in our RT model inks it to the problem of the growth of bacterial colonies. Recently, the authors of \cite{narla_traveling-wave_2021}  
developed a growth-expansion model which generates traveling waves in bacterial che\-motaxis, in accord with experimental observations. We show that traveling waves also 
arise in our much simpler 1d run-and-tumble model.

The paper is organized as follows. In Section 2, we introduce our RT-model as a toy model, inspired by the cell cycle of {\it CC}. In Section 3, we first focus on the case of free cells 
for which we derive the conditions for stability of the system when spatial dependencies are neglected. In Section 4 we consider the spatial dependence built into the model 
and study the mean displacement (MD) and mean squared displacement (MSD) for a single cell in the process of duplicating, both showing a surprising $t^3$ regime for short times.
Allowing the cells to interact via both attraction and repulsion mechanisms, this antagonistic effect is found to lead to structure formation: we numerically find traveling wave solutions
of the system density and quantify their occurrence in a non-equilibrium state diagram. Finally we discuss how the model performs with parameter values specific of {\it CC}. 
Section 5 concludes the paper with a discussion of the results of our model and a brief outlook on further work.

\section{The model}

Inspired by the reproductive behavior of {\it Caulobacter crescentus} we consider a one-dimensional toy model representing bacteria that can actively move rightward, leftward or settle down, 
and that when settled double in number. We note, that {\it CC} performs a run-reverse-flick motion~\cite{grognot_more_2021}, where the bacterium first performs a forward motion, then reverses its 
direction of motion and in a third step makes a turn mediated by a buckling instability in its flagellum~\cite{son_bacteria_2013}. Since our setup is one dimensional, the run-reverse-flick motion is equivalent to a run and tumble motion.

\begin{figure}
\begin{center}
\begin{tikzpicture}[scale=1.5]
  \draw[blue,rounded corners=10,thick]
     (0,0) rectangle (1,1)
      node[pos=.5] {$\rho_-$} ;
    \draw [orange,-{Stealth}](.95,1.05) -- (1.45,1.55) node[midway,above left] {$\lambda_s$};
  \draw[olive,rounded corners=10,thick,shift={(1.5,1.5)}]
    (1,0) rectangle (0,1)
     node[pos=.5] {$\rho_0$} ;
     \draw [teal,-](2,1.2) -- (2,1.45) node[pos=0,below] {$\lambda_d$};
     \draw [teal,{Stealth}-](1.05,.95) -- (2,1.2) node[midway,below right] {};
     \draw [teal,{Stealth}-](2.95,.95) -- (2,1.2) node[midway,below right] {};
     \draw [orange,{Stealth}-](2.55,1.55) -- (3.05,1.05) node[midway,above right] {$\lambda_s$};
  \draw[red,rounded corners=10,thick,shift={(3,0)}]
    (1,0) rectangle (0,1)
     node[pos=.5] {$\rho_+$} ;
     \draw [violet,{Stealth}-{Stealth}](1,0.5) -- (3,0.5) node[midway,below] {$\lambda_e$};
  	 \draw [black,-{Stealth}](0,1) -- (-.5,1.5) node[midway,below left] {$\mu$};
  	 \draw [black,-{Stealth}](4,1) -- (4.5,1.5) node[midway,below right] {$\mu$};
  	 
    \draw [blue,-{Stealth}](.5,.25) -- (.25,.25) node[midway,below right] {$v_-$};
    \draw [red,-{Stealth}](3.5,.25) -- (3.75,.25) node[midway,below] {$v_+$};
    \foreach \Point in {(.5,.25), (2,1.75), (3.5,.25)}{\node at \Point {\textbullet};}
\end{tikzpicture}
\end{center}
\caption{Graphical representation of the transition rates among different species. These transitions are motivated by the cell cycle of {\it Caulobacter crescentus}, 
that either moves actively or settles down to reproduce. Our model contains three different species: the cells moving to the right $\rho_+$, those moving to the left $\rho_-$ 
and the settled ones $\rho_0$. The moving cells can either settle via the rate $\lambda_s$, move in the opposite direction with $\lambda_e$ or die with $\mu$. 
Settled cells duplicate via $\lambda_d$, and generate both a left- and a right-moving cell.}
\label{Fig1}
\end{figure}
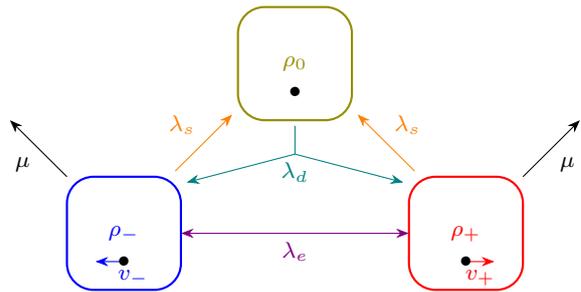

The `cell cycle' of our three-state RT model motivated by {\it CC} is summarized in Figure \ref{Fig1}. We allow for three populations with 
the number densities $\rho_+(x,t)$, $\rho_-(x,t)$ and $\rho_0(x,t)$, functions of space $x$ and time $t$, respectively corresponding to right and left movers, 
and to the sedentary population. The `cell cycle' step is given by the rate of settling down, $\lambda_s$,
which can occur from either moving state, and the cell doubling with rate $\lambda_d$ with which a sedentary bacterium gives rise to a pair of right- and left-moving cells.
The exchange of direction, i.e. the RT step, is denoted by $\lambda_e$. Finally, $\mu$ is the death rate, which we consider for motile cells only. In a proliferating system,
this rate prevents exponential growth. 
This idealized {\it CC}-`cell cycle' is implemented in terms of evolution equations for the cell number densities. In the case where there is no death or proliferation,
the number densities can also be interpreted as probability densities and the evolution equations correspond to Fokker-Planck equations.

As the bacteria are micron-sized swimmers, we assume a low Reynolds number and overdamped dynamics.
To describe this behavior mathematically we first group the three densities into the vector of densities $\boldsymbol\rho = (\rho_+,\rho_0,\rho_-)$.
The dynamics of the system will then be described by the differential equation
\begin{equation} \label{gee} 
\partial_t \boldsymbol\rho =  {\mathcal D} \partial_x^2 \boldsymbol\rho + \partial_x [(\partial_x {\mathcal U}) \cdot \boldsymbol\rho] - {\cal V} \cdot \partial_x \boldsymbol\rho + 
{\cal M}\boldsymbol\rho
\end{equation}
which generalizes the standard expression of growth-expansion equations of logistic growth, usually formulated for a single density \cite{narla_traveling-wave_2021}. In 
Eq. (\ref{gee}), the first term is a diffusion term where the matrix ${\cal D}$ has the form
\begin{equation}
\mathcal{D} = 
\begin{pmatrix}
D & 0 & 0 \\
0 & 0 &  0 \\
0 & 0 &  D 
\end{pmatrix}
\end{equation} 
since the sedentary particles do not diffuse. The second term on the right-hand side is a nonlinear diffusion coefficient 
containing an interaction matrix ${\cal U}$ of the form
\begin{equation}
\mathcal{U} = 
\begin{pmatrix}
-\kappa \rho_0 & 0 & 0 \\
0 & \kappa_0 \rho_0  &  0 \\
0 & 0 &  -\kappa \rho_0
\end{pmatrix}\, .
\end{equation} 
The matrix entries describe attractive interactions (negative sign) of the moving cells to regions in which particles have settled and repulsive interactions among settled cells
(positive sign) in order to mimic biofilm behaviour. 
The third term on the right-hand side of Eq. (\ref{gee}) describes the active motion of the particles in the right and left directions along the line. Hence
\begin{equation}
\mathcal{V} = 
\begin{pmatrix}
v_+ & 0 & 0 \\
0 & 0 &  0 \\
0 & 0 &  -v_-  
\end{pmatrix}\, .
\end{equation} 
Finally, we have for the cell cycle or population dynamics part, following the transitions shown in Figure 1, the matrix ${\cal M}$ given by 
\begin{equation}
\mathcal{M} = 
\begin{pmatrix}
-(\lambda_s+\lambda_e+\mu) & \lambda_d & \lambda_e \\
\lambda_s & -\lambda_d & \lambda_s \\
\lambda_e & \lambda_d & -(\lambda_s+\lambda_e+\mu) 
\end{pmatrix}\, .
\end{equation}
Given that our run-and-tumble model allows for proliferation and death of cells, it is important to recognize that the population dynamics
of Eq. (\ref{gee}) is the linear limit of the more general nonlinear decay-growth equation
\begin{eqnarray} \label{gee3} 
\partial_t \boldsymbol{\rho} &=& {\mathcal D} \partial_x^2 \boldsymbol\rho + \partial_x [(\partial_x {\mathcal U}) \cdot \boldsymbol\rho] - {\cal V} \cdot \partial_x \boldsymbol\rho \nonumber\\
& &+\mathcal{M}_D \boldsymbol{\rho} +  \mathcal{M}_{OD} \boldsymbol{\mathcal{R}}(\boldsymbol{\rho}).
\end{eqnarray}
 In Eq. (\ref{gee3}), $\mathcal{M}_D $ and $ \mathcal{M}_{OD} $ are the diagonal and off-diagonal parts of the matrix $\mathcal{M}$, i.e., one has
$\mathcal{M} = \mathcal{M}_D  + \mathcal{M}_{OD} $. The diagonal part describes the cell number decay, while the off-diagonal part describes the growth of the 
cell population. In order to limit growth, the non-diagonal term is generally nonlinear and saturating at the carrying capacity, as is common in 
growth-expansion models, see, e.g. \cite{narla_traveling-wave_2021}. The vector $\mathcal{R} $ is thus given by 
\begin{eqnarray} 
\boldsymbol{\mathcal{R}} = 
\begin{pmatrix}
\rho_+(1 - \frac{\rho_+}{\rho_{+,c}})\\
\rho_0(1 - \frac{\rho_0}{\rho_{0,c}})\\
\rho_-(1 - \frac{\rho_-}{\rho_{-,c}}) 
\end{pmatrix}\, . \nonumber
\end{eqnarray}
where the carrying capacity is given by the vector 
\begin{equation}
{\boldsymbol\rho}_c(x) \equiv \left(\rho_{+,c}(x),\rho_{0,c}x),\rho_{-,c}(x)\right)\, .
\end{equation}
The linear limit of Eq. (\ref{gee3}) is reached for $ |\boldsymbol{\rho}| \ll |\boldsymbol{\rho}_c| $. It is important to notice that since $\mathcal{R}$ is only applied to one part of the $\mathcal{M}$ matrix, the stationary values reached by the population in the linear limit will not necessarily be those given by $\boldsymbol\rho_c$. The main benefit of the nonlinear model is that it prevents the number of cells from exploding independently of the parameters. In this manuscript we will mainly focus on the linear case, while explicitly referring to the full nonlinear growth equation if needed.

\section{Free cells}
We start by setting the cell interaction parameters $\kappa=\kappa_0=0$, and hence consider free cells.

\subsection{Population dynamics}

\begin{figure}[!htbp]
 \begin{center}
 \includegraphics[width=.45\textwidth]{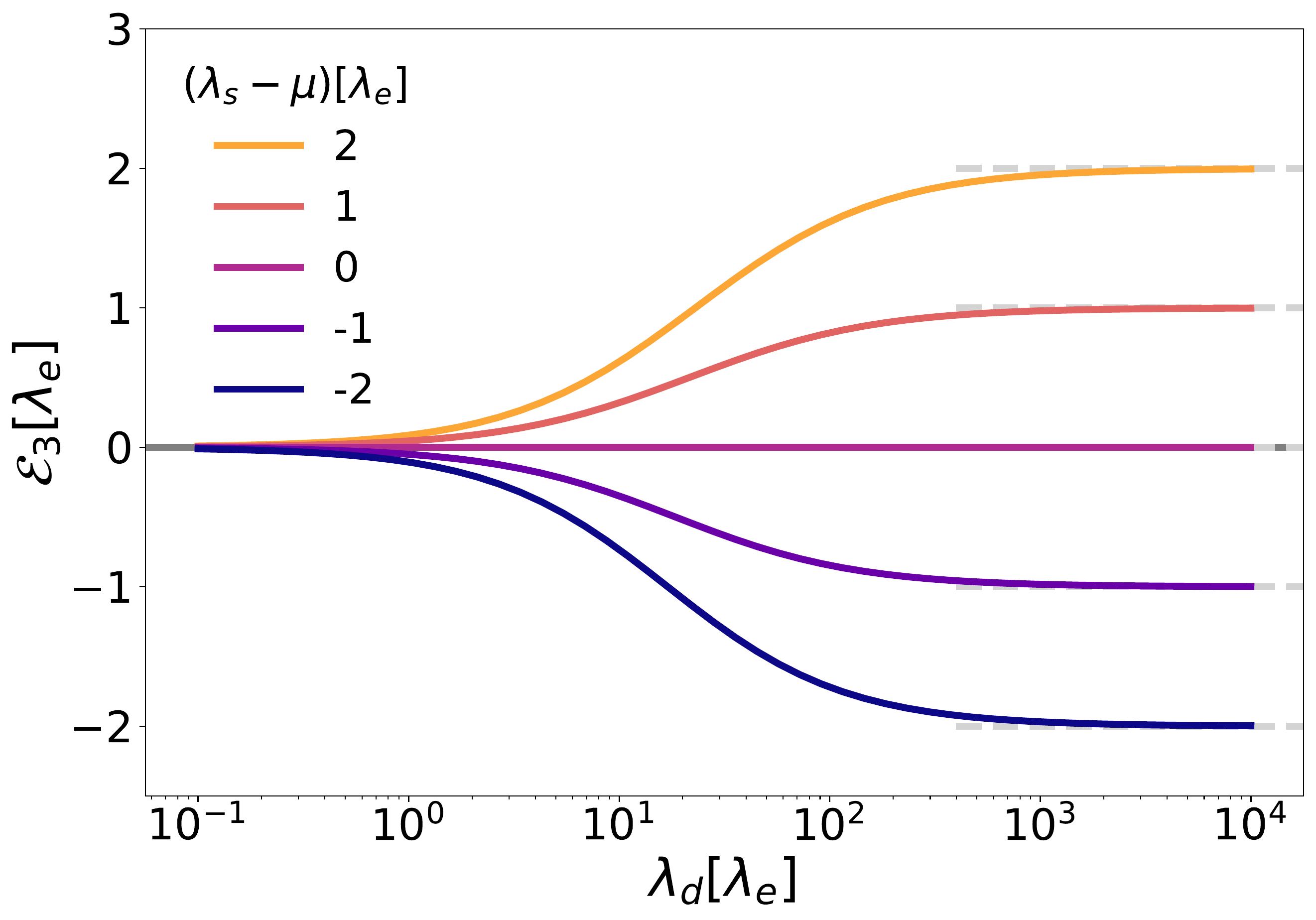}
 \caption{Unstable eigenvalue $\mathcal{E}_{3}$ (solid lines) as a function of doubling rate $\lambda_d$ for different values of $\lambda_s$ (color code) and $\mu=10\lambda_e$. 
 The sign of $\mathcal{E}_{3}$ is the same of $\lambda_s-\mu$, and its value also stabilizes at $\lambda_s-\mu$  for very large values of $\lambda_d$ (dashed  lines).}
\label{Fig2}
\end{center}
\end{figure}

\begin{figure}[!htbp]
 \begin{center}
 \includegraphics[width=.45\textwidth]{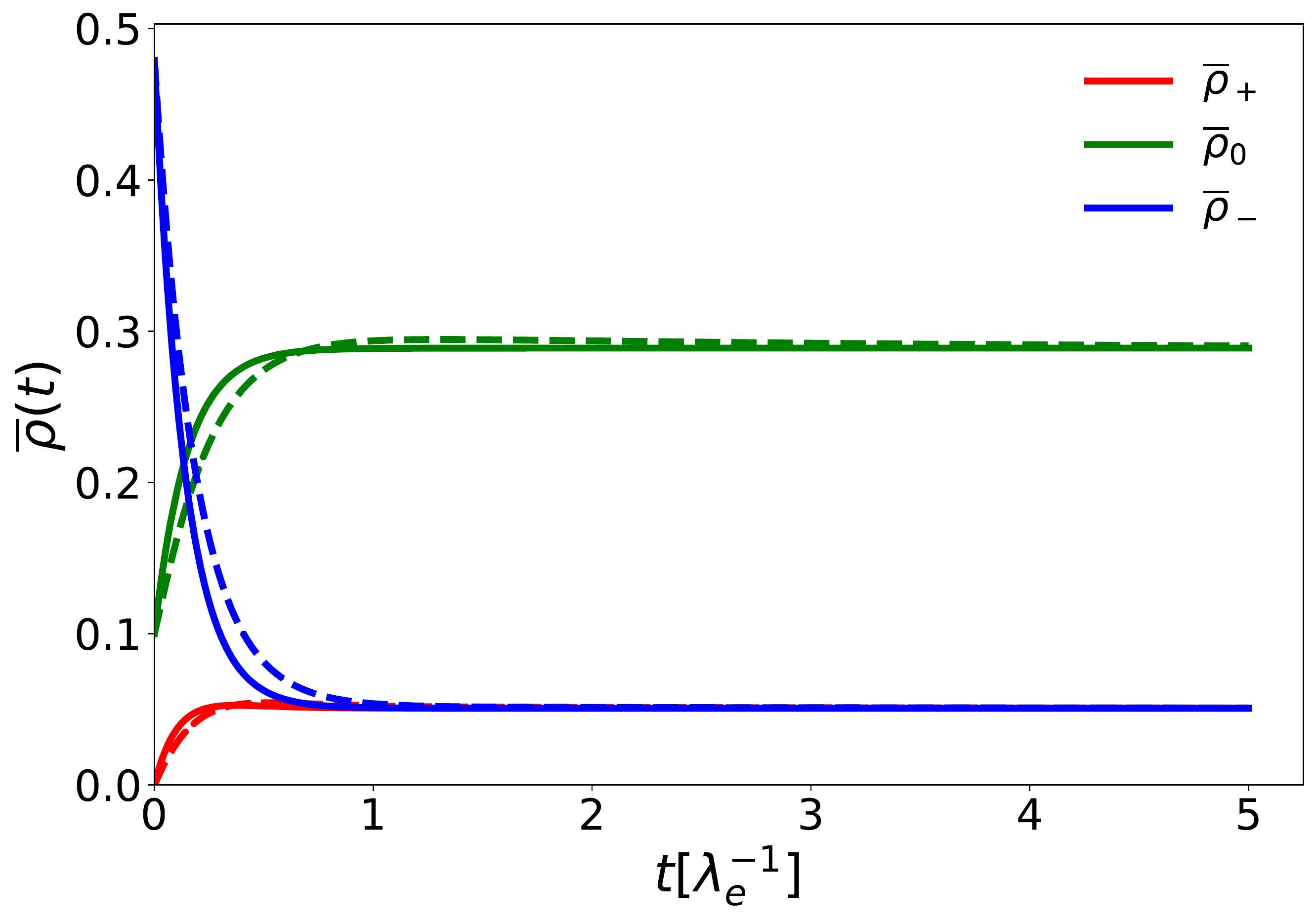}
 \caption{Space averages of right- $\overline{\rho}_+$, left-moving $\overline{\rho}_-$ and sedentary $\overline{\rho}_0$ cells as functions of time, both for the linear model (solid lines) with $\lambda_s=\mu$  and for the nonlinear model (dashed lines) with $\lambda_s=3\mu$. For both models $\lambda_d$ is set to be equal to $\lambda_e$, while $\mu=\lambda_e$ in the nonlinear model and $\mu=2.848\lambda_e$ in the linear one. As initial conditions we chose the constant values $\boldsymbol\rho(x,0)=(0,.1,0.479)$ for both models. For the nonlinear model we further set the carrying capacity $\boldsymbol\rho_c(x)=(1,1,1)$.}
\label{Fig3}
\end{center}
\end{figure}

In this section we further set $D=0$ as well as the velocities $v_+ = v_- =0 $, thus we first study free cells undergoing the pure population dynamics given by
\begin{equation} \label{gee2} 
\partial_t \boldsymbol\rho(x,t) = {\cal M}\boldsymbol\rho(x,t)\, .
\end{equation}
This linear system of equations can be solved analytically via matrix calculations, leading to
\begin{equation}
\label{4}
\boldsymbol\rho(x,t)=\text{e}^{\mathcal{M}t}\boldsymbol\rho(x,0)=\mathcal{P}\text{e}^{\mathcal{E}t}\mathcal{P}^{-1}\boldsymbol\rho(x,0).
\end{equation}
$\mathcal{P}$ is the eigenvector matrix of $\mathcal{M}$ and $\mathcal{E}$ is the diagonal matrix containing the eigenvalues of $\mathcal{M}$, that are
\begin{align}
\mathcal{E}_{1}=&-(\mu+2\lambda_e+\lambda_s) \nonumber \\
\mathcal{E}_{2}=&-(\mu+\lambda_d+\lambda_s+\Lambda)/2 \label{5}\ \\
\mathcal{E}_{3}=&-(\mu+\lambda_d+\lambda_s-\Lambda)/2, \nonumber 
\end{align}
where $\Lambda=\sqrt{(\mu+\lambda_d+\lambda_s)^2+4\lambda_d(\lambda_s-\mu)}$.
We notice that the first two eigenvalues are always negative and therefore stable, while the sign of the third depends on $\lambda_s-\mu$, which can become unstable. 
This instability facilitates an exponential growth of the colony. In fact, for small values of $\lambda_d(\lambda_s-\mu)$ with respect to $\mu+\lambda_d+\lambda_s$ 
the unstable eigenvalue becomes
\begin{equation}
\label{6}
\mathcal{E}_{3}\simeq \frac{\lambda_d(\lambda_s-\mu)}{\mu+\lambda_d+\lambda_s}.\\
\end{equation}
The exponential growth or collapse of the system is therefore decided by the difference of $\lambda_s$ and $\mu$, or in different terms, the separating line between the two behaviors is $\lambda_s=\mu$. It is also worth pointing out that in the case of instant doubling, that is the limit of $\lambda_d\rightarrow \infty$, $\mathcal{E}_{3}$ simply reduces to $\lambda_s-\mu$, as can be seen in Figure \ref{Fig2}. Physically this is expected, as in this model cells can double only when settled and can die only when moving, meaning that the growth or decay of the system size depends exclusively on whether a moving cell is faster in settling or dying. 
In the case of $\lambda_s=\mu$, it is possible to calculate the stationary value of of $\boldsymbol\rho(x,t\rightarrow \infty)$ as a 
function of the initial conditions $\boldsymbol\rho(x,0)$:
\begin{align}
\label{7}
\rho_+(x,t\rightarrow \infty)&=\frac{\lambda_d}{2(2\mu+\lambda_d)}R(x,0) \nonumber \\
\rho_0(x,t\rightarrow \infty)&=\frac{\mu}{2\mu+\lambda_d}R(x,0)  \\
\rho_-(x,t\rightarrow \infty)&=\frac{\lambda_d}{2(2\mu+\lambda_d)}R(x,0), \nonumber 
\end{align}
where $R(x,0)=2\rho_0(x,0)+\rho_-(x,0)+\rho_+(x,0)$.
Since the exchange rate between right $\rho_+$ and left $\rho_-$  moving cells is symmetric, the amounts of left and right moving cells are the same in the stationary state ($\rho_+=\rho_-$, 
see also Figure \ref{Fig3}).
Furthermore, if $\lambda_d=2\mu=2\lambda_s$ all the three populations equilibrate to the same value, independently of the initial conditions. In the case of $\lambda_s>\mu$ it is always possible in the frame of the nonlinear growth model to find values of $\boldsymbol\rho_c$ for which the populations stabilize around the values given by Eq.(\ref{7}).
Figure \ref{Fig3} shows the linear and nonlinear model equations with different parameters and with the same stationary values.

\subsection{Density dynamics}
\label{Density}
We now set the running speeds $v_\pm$ and the diffusion constant $D$ to finite values, in order to study the evolution of spatial quantities of the system, such as the mean displacement MD $ = \langle x-x_0\rangle$, the mean-squared displacement MSD $ =\langle (x-x_0)^2\rangle$ and all the higher order moments, where $x_0$ is the average position of the system at $t=0$. Here, the average 
$\langle (\cdot) \rangle$ is defined as $\int_{-\infty}^\infty (\cdot)P(x,t)\text{d}x$, where the total probability $P(x,t)$ is
\begin{equation}
P(x,t) \equiv \frac{1}{N(t)}(\rho_+(x,t)+\rho_0(x,t)+\rho_-(x,t)), 
\end{equation}
$N(t)\equiv N_0(t)+N_+(t)+N_-(t)$ is the total number of cells, $N_\alpha(t)=\int_{-\infty}^\infty \rho_\alpha(x,t)\text{d}x$ is the number of cells in phase $\alpha$ and $\alpha$ can be $(+,-,0)$.

In order to compute averages, we first solve the system by using a Fourier transform ($FT$):
\begin{equation}
\label{8}
\dot{\tilde{\boldsymbol\rho}}(k,t) = \left(-k^2\mathcal{D} -\text{i}k\mathcal{V} +\mathcal{M}\right)\tilde{\boldsymbol\rho}(k,t),
\end{equation}
where $\tilde{\boldsymbol\rho}(k,t)=FT(\boldsymbol\rho(x,t))$ is the Fourier transform of $\boldsymbol\rho(x,t)$ and $k$ is the wave number conjugate to $x$.
Similarly to the constant density case, the solution in Fourier space will be given by
\begin{equation}
\label{10}
\tilde{\boldsymbol\rho}(k,t)=\text{exp}\left[(-k^2\mathcal{D} -\text{i}k\mathcal{V} +\mathcal{M})t\right]\tilde{\boldsymbol\rho}(k,0).
\end{equation}
One can use the solution of this equation to extract the intermediate scattering function (\textit{ISF})
\begin{equation}
\mathcal{F}(k,t)\equiv \tilde{P}(k,t)\tilde{P}(-k,0)N(t).
\end{equation}
The \textit{ISF}  can be related to the different moments of the density \cite{kurzthaler_intermediate_2016} by differentiation:
\begin{equation}
\label{11}
\langle (x(t)-x_0)^n \rangle =\left.\frac{\text{i}^n}{N(t)}\frac{\partial^n}{\partial k^n}\mathcal{F}(k,t)\right|_{k=0},
\end{equation} 
valid in one dimension (see Appendix).
\\

We can also define an average for each cell population and the relative \textit{ISF}:
\begin{eqnarray}
\langle (\cdot) \rangle_\alpha &\equiv & \int_{-\infty}^\infty (\cdot)\frac{\rho_\alpha(x,t)}{N_\alpha(t)}\text{d}x,\\
\mathcal{F}_\alpha(k,t)&\equiv & \frac{\tilde{\rho}_\alpha(k,t)\tilde{\rho}_\alpha(-k,0)}{N_\alpha(0)}.
\end{eqnarray}
The expression corresponding to Eq.(\ref{11}) is then given by
\begin{equation}
\label{11a}
\langle (x(t)-x_0)^n \rangle_\alpha =\left.\frac{\text{i}^n}{N_\alpha(t)}\frac{\partial^n}{\partial k^n}\mathcal{F}_\alpha(k,t)\right|_{k=0}.
\end{equation}
First we will discuss the behavior of the whole distribution $P(x,t)$.
For simplicity, we will consider the initial condition $\boldsymbol\rho(x,0) = (0,\delta(x),0)$ which is physically relevant, as it describes a cell initially settled in $x=0$ in the process of reproducing. 
We do not focus on the case of an initially mobile cell, as the short-time behaviours of both the MD and MSD turn out to be simply linear and the long-time behaviours are identical to that 
of the initially settled cell case. We further remark that our analysis does not assume the condition $\mu=\lambda_s$ for a stable population in the linear growth model.

\subsubsection{Full distribution}

\begin{figure*}[!htbp]
 \begin{center}
 \includegraphics[width=.9\textwidth]{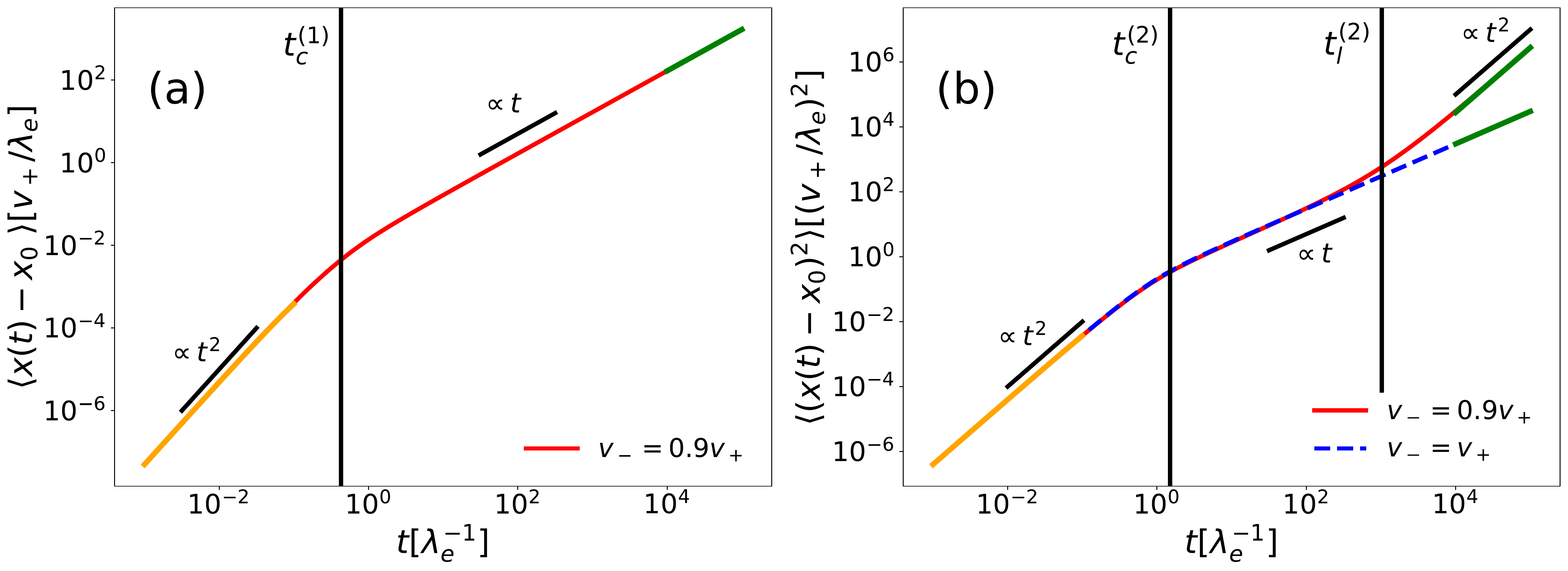}
 \caption{(a) Mean displacement (MD), (b) mean-squared displacement (MSD), respective crossing times $t_c^{(1)}$, $t_c^{(2)}$ and short- and long-time approximations for the initial conditions $\boldsymbol\rho(x,0)=(0,\delta(x),0)\lambda_e/v_+$, all rates equal to $\lambda_e$ and $D=0.2v_+^2/\lambda_e$. In (b) the solid red line shows unequal swim velocities ($v_-=0.9v_+$) and the dashed blue line equal swim speeds ($v_-=v_+$). The orange lines represent the short-time approximations, while the green lines are the long-time approximations.}
\label{Fig4}
\end{center}
\end{figure*}

When $v_+\neq v_-$, the MD is non-zero and we observe two different regimes: for short times it grows as $t^2$, while for long times it is proportional to $t$, as shown in Figure \ref{Fig4}(a). 
The short-time expansion of the MD in fact yields 
\begin{eqnarray}
\label{12}
\langle x(t)-x_0\rangle &=&\lambda_d v_dt^2\nonumber\\
 & & -\frac{1}{3}\lambda_d v_d(2\mu +4\lambda_d +\lambda_s)t^3\nonumber\\
 & &+\mathcal{O}\left(t^4\right),
\end{eqnarray}
where $v_d=(v_+-v_-)/2$. The expression shows that both the transition rates and the running speeds have a role in determining this initial scaling regime. 
This can be interpreted as a composition of the doubling mechanism and the system acceleration given by cells suddenly starting to move. We can further define 
the typical crossover time  $t_c^{(1)}$ as the ratio between absolute values of the coefficients of the $t^2$ and $t^3$ scalings, as this is the time at which the the $t^2$ order contribution becomes smaller than the following ones \cite{breoni_active_2020, breoni_active_2021}. This is a good estimate of the average time at which the dynamics is not 
dominated by the initial doubling anymore:
\begin{equation}
\label{13}
t_c^{(1)}=\frac{3}{2\mu +4\lambda_d +\lambda_s}.
\end{equation}
The long-time expansion of the MD yields
\begin{equation}
\label{13a}
\langle x(t)-x_0\rangle = \frac{4 v_d\lambda_d\lambda_s}{\Lambda(\mu-\lambda_d+\lambda_s+\Lambda)}t+ \mathcal{O}\left(t^0\right),
\end{equation}
where $\Lambda$ is the same of Eq.(\ref{5}).

\begin{figure*}[!htbp]
 \begin{center}
 \includegraphics[width=.8\textwidth]{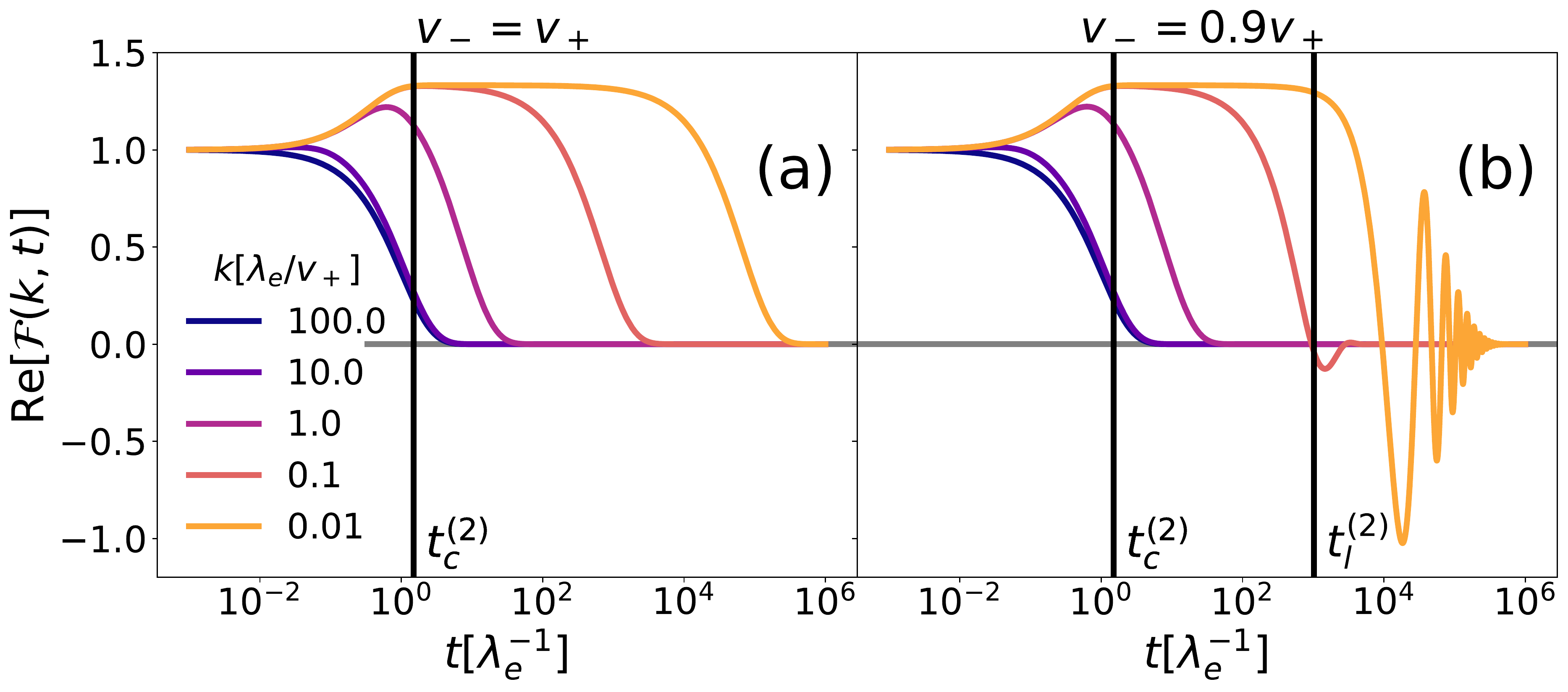}
 \caption{Real part of the intermediate scattering function $\mathcal{F}(k,t)$ for (a) equal swimming speeds and (b) unequal swimming speeds ($v_-=0.9v_+$) for the initial conditions $\boldsymbol\rho(x,0)=(0,\delta(x),0))\lambda_e/v_+$, all rates equal to $\lambda_e$ and $D=0.2v_+^2/\lambda_e$. The black lines represent the MSD short crossing time $t_c^{(2)}$ and, in the case of different speeds, long crossing time $t_l^{(2)}$.}
\label{Fig5}
\end{center}
\end{figure*}

As far as the MSD is concerned, in Figure \ref{Fig4}(b) we still see a $t^2$ regime for short times, while the long-time behavior depends on the difference between 
$v_-$ and $v_+$. In the case they are the same, we will only see a diffusive long-time regime while otherwise this diffusive regime transitions into a ballistic one. 
The smaller the difference between the running speeds, the longer is the time to reach the ballistic regime.
We further calculate the short-time expansion of the MSD:
\begin{eqnarray}
\label{14}
\langle (x(t)-x_0)^2\rangle &=& 2D\lambda_d t^2\nonumber\\
 & & -\frac{2}{3}\lambda_d\left(D(2\mu +4\lambda_d +\lambda_s)-v_a^2\right)t^3\nonumber\\
 & &+\mathcal{O}\left(t^4\right),
\end{eqnarray}
where $v_a=\sqrt{(v_+^2 +v_-^2)/2}$.
Again, we define a crossing time $t_c^{(2)}$  for the MSD as the ratio between the absolute values of the coefficients of the $t^2$ and $t^3$ scalings:
\begin{equation}
\label{15}
t_c^{(2)}=\frac{3D}{\left|D(2\mu +4\lambda_d +\lambda_s)-v_a^2\right|}.
\end{equation}
If we change the population rates we observe that the growth or decay in the number of cells does not influence qualitatively the scalings we just described for both the MD and MSD. The formula for the long-time expansion of the MSD and the relative crossing time $t_l^{(2)}$ between the long-time regimes $\propto t$ and $\propto t^2$ are quite involved, so we refrain from showing them here.
\\
Finally, we study directly the full intermediate scattering function $\mathcal{F}(k,t)$ as it carries more information than the MSD and MD. In Figure \ref{Fig5} (a), that is in the case of equal velocities, we can see that the real part of $\mathcal{F}(k,t)$ that generates the MSD among all other even moments, decays rapidly for small length scales (i.e. large $k$) while it has three distinct regimes for large length scales. At first the function decays or grows, following the growth in the number of cells, then at time $t^{(2)}_c$ it plateaus for a time that grows larger as $k$ gets smaller, and finally decays completely. 
The plateau, starting after the transition of the cell to its moving stage at time $t^{(2)}_c$, is generated by the active cells going back to the settled stage and not moving anymore, while the final decay represents the long-time diffusive behavior that we have already seen in the MSD. In Figure \ref{Fig5} (b) we see how unequal velocities change the intermediate scattering function by introducing an oscillating behavior at long times. This is a signature of ballistic motion and of a non-vanishing imaginary part of $\mathcal{F}(k,t)$ that generates the odd moments like the MD.
\\

\subsubsection{Distribution of settled cells}

\begin{figure*}[!htbp]
 \begin{center}
 \includegraphics[width=.9\textwidth]{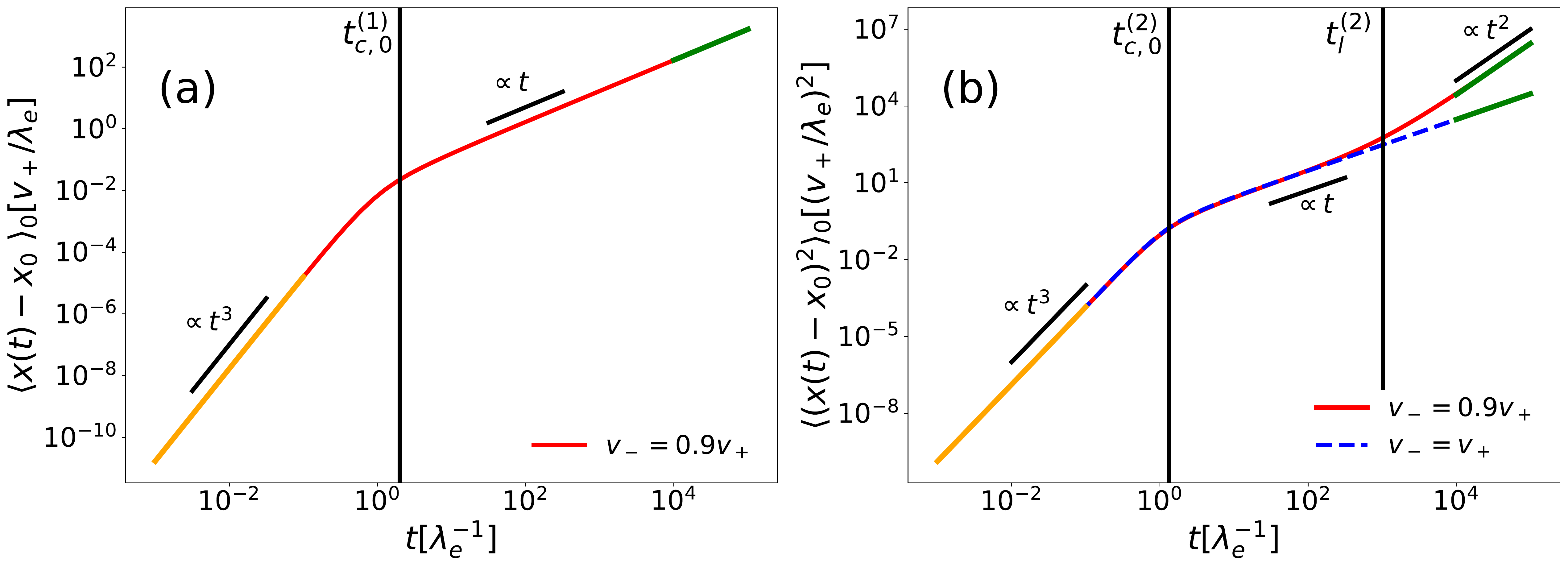}
 \caption{(a) Mean displacement (MD), (b) mean-square-displacement (MSD), respective crossing times $t_c^{(1)}$, $t_c^{(2)}$ and short- and long-time approximations for settled cells, with initial conditions $\boldsymbol\rho(x,0)=(0,\delta(x),0))\lambda_e/v_+$, all rates equal to $\lambda_e$ and $D=0.2v_+^2/\lambda_e$. In (b) the solid red line shows unequal swim velocities ($v_-=0.9v_+$) and the dashed blue line equal swim speeds ($v_-=v_+$). The orange lines represent the short-time approximations, while the green lines are the long-time approximations.}
\label{Fig6}
\end{center}
\end{figure*}

The main feature of the MD and MSD of the settled cells is that they both show an initial $t^3$ regime, as shown in Figure \ref{Fig6}. The short time expansion of the MD is
given by
\begin{eqnarray}
\langle x(t)-x_0\rangle_0 &=&\frac{1}{3}\lambda_s\lambda_dv_dt^3\nonumber\\
 & & -\frac{1}{6}\lambda_s\lambda_dv_d(\mu -\lambda_d +\lambda_s)t^4\nonumber\\
 & &+\mathcal{O}\left(t^5\right),
\end{eqnarray}
with the crossing time between the $t^3$ and $t^4$ regimes $t_{c,0}^{(1)}$ being:
\begin{equation}
t_{c,0}^{(1)}=\frac{2}{|\mu -\lambda_d +\lambda_s|}.
\end{equation}
The MSD shows the initial $t^3$ regime as well:
\begin{eqnarray}
\langle (x(t)-x_0)^2\rangle_0 &=&\frac{2}{3}D\lambda_s\lambda_d t^3\nonumber\\
 & & -\frac{1}{6}\lambda_s\lambda_d\left(2D(\mu -\lambda_d +\lambda_s)-v_a^2\right)t^4\nonumber\\
 & &+\mathcal{O}\left(t^5\right).
\end{eqnarray}
with the crossing time $t_{c,0}^{(2)}$:
\begin{equation}
t_{c,0}^{(2)}=\frac{4D}{\left| 2D(\mu -\lambda_d +\lambda_s)-v_a^2\right|}.
\end{equation}
The reason why we observe the $t^3$-behaviour for short times is the fact that the settled population can only change by doubling, moving and then settling, with each one of 
these processes being at least of order $t$. We also notice that for $D=0$ the MSD grows initially with $t^4$, as in this case the short time MSD for moving cells grows with $t^2$ and not $t$.\\
The long-time asymptotes for both MD and MSD of the settled particles are identical to those of the whole population.

\section{Interacting cells}

\subsection{Attraction to settled regions}

We now discuss the case of interacting cells. Our model contains an effective attractive force that pushes the moving cells towards the regions where the density of settled cells is larger. This force is meant to represent how bacteria tend to assemble in resource-rich regions to reproduce or how they accumulate in order to form biofilms \cite{hall-stoodley_bacterial_2004,mazza_physics_2016}; therefore the parameter $\kappa >  0$ in Eq.(\ref{gee}). The interaction terms $\kappa\partial_x(\partial_x(\rho_0)\rho_\pm)$ render the equation nonlinear such that it is not analytically solvable. Instead we first 
perform a linear stability analysis around the homogeneous stationary solution to the linear system $\hat{\boldsymbol\rho}$ computed in Eq. \eqref{7} (see also Fig.~\ref{Fig3}) by adding a small 
perturbation $\delta \boldsymbol\rho(x,t)$ and neglecting the nonlinear terms in the perturbation $(\delta \boldsymbol\rho(x,t))^2$. We then arrive at the following system of equations for the perturbation:
\begin{align}
\partial_t\delta\rho_+=&-v_+\partial_x\delta\rho_+-\kappa\partial_x^2(\delta\rho_0)\hat{\rho}_++D\partial_x^2\delta\rho_+ \nonumber \\
&-(\lambda_s+\lambda_e+\mu)\delta\rho_++\lambda_e\delta\rho_-+\lambda_d\delta\rho_0 \nonumber \\
\partial_t\delta\rho_0=&-\lambda_d\delta\rho_0+\lambda_s(\delta\rho_++\delta\rho_-) \label{18}  \\
\partial_t\delta\rho_-=&v_-\partial_x\delta\rho_--\kappa\partial_x^2(\delta\rho_0)\hat{\rho}_- +D\partial_x^2\delta\rho_- \nonumber \\
&-(\lambda_s+\lambda_e+\mu)\delta\rho_-+\lambda_e\delta\rho_++\lambda_d\delta\rho_0, \nonumber 
\end{align}
where the stationary values for the density are symmetric, $\hat{\rho}_+=\hat{\rho}_-$. 
We apply both a Fourier transform in space and a Laplace transform in time to Eq.~\eqref{18} and solve the resulting characteristic equation of the system. 
We obtain three different solutions for the eigenvalues of the system $s_i(k)$, of which only one, $s_1(k)$, can have a positive real part. In the following we focus 
on $s_1(k)$, since its positive real part introduces instabilities in the system.
\\

First of all, for $k\rightarrow 0$, the value of $s_1(k)$ is one of the eigenvalues of the system matrix where the initial densities are constant, and more specifically the one that can be positive:
\begin{equation}
\label{19}
s_1(0)=\mathcal{E}_{3}\simeq\frac{\lambda_d(\lambda_s-\mu)}{\mu+\lambda_d+\lambda_s}.
\end{equation}
This means that one of the conditions for the system to be stable is that the number of cells does not grow exponentially, which is expected.
\\

\begin{figure}[!htbp]
 \begin{center}
 \includegraphics[width=.45\textwidth]{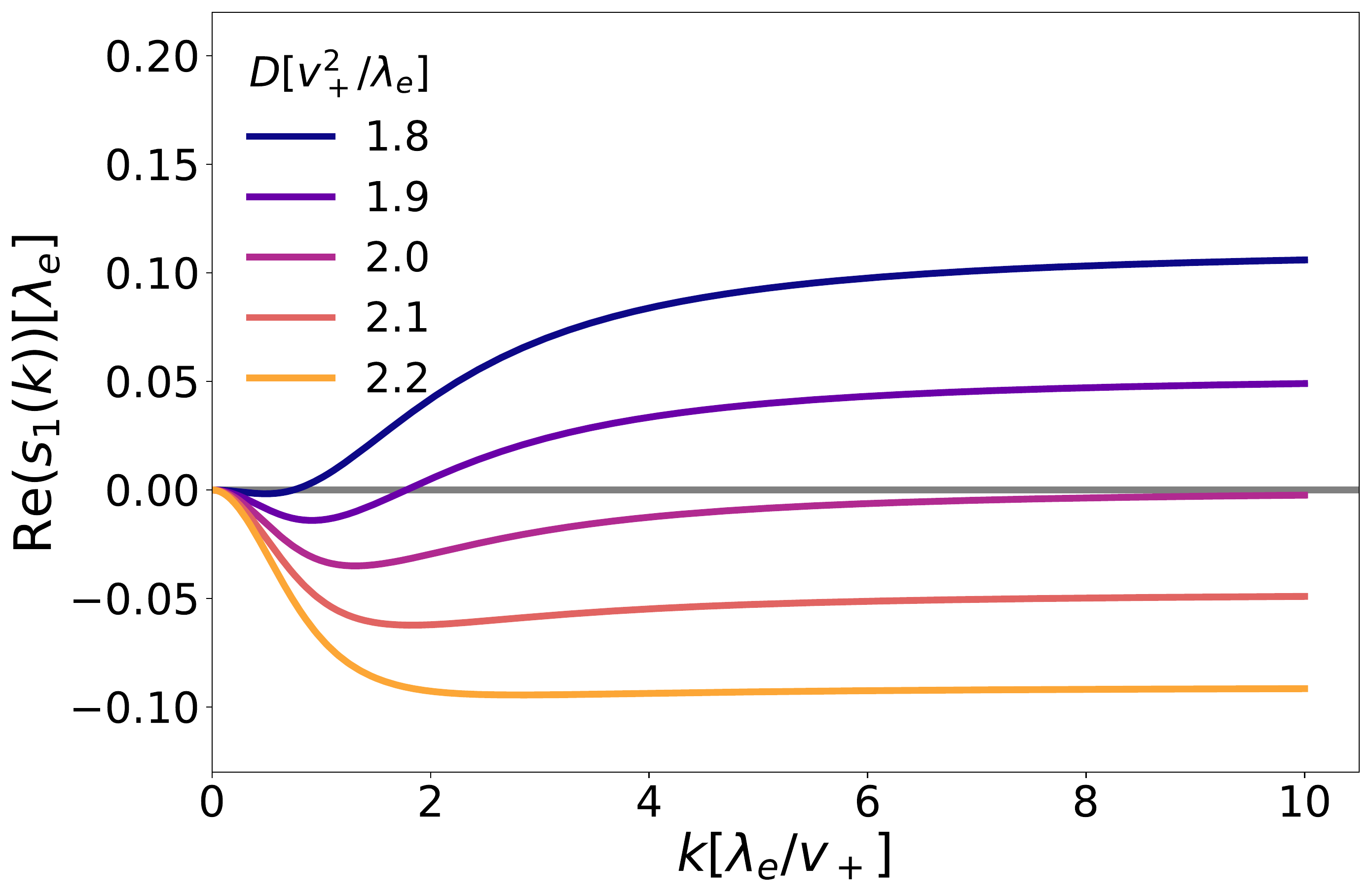}
 \caption{Eigenvalue $s_1(k)$ as a function of wavenumber $k$ for different values of $D$, where all rates are equal to $\lambda_e$, $v_-=v_+$ and $\kappa=\lambda_e^{-1}$. }
\label{Fig7}
\end{center}
\end{figure}

The second limit we consider is $k\rightarrow \infty$. We have that
\begin{equation}
\label{20}
\lim_{k\rightarrow \infty}s_1(k)\rightarrow \frac{2\kappa \hat{\rho}_+ \lambda_s}{D}-\lambda_d.
\end{equation}
This second condition states that the diffusion constant contrasts directly the instabilities generated by a large settling rate and the attractive constant $\kappa$, as it disperses too large clusters of active cells, while a large doubling rate helps the stability by reducing the size of groups of settled cells. Knowing the limits of $s_1(k)$ in $k=0$ and $k=\infty$, i.e long- and short-range perturbations respectively, we are sure that the system will be unstable if the real part of either of them is larger than zero, giving us two stability conditions for the system:
\begin{align}
&\mu\geq\lambda_s, \nonumber \\
&\lambda_d\geq\frac{2\kappa \hat{\rho}_+ \lambda_s}{D}. \label{stability}
\end{align}
 For $D=0$, $s_1(k)$ grows asymptotically like $k$, making the system always unstable. In Figure \ref{Fig7} we show the behavior of the eigenvalue Re$(s_1(k))$ for different values of $D$. Notice that for the set of parameters considered, if $D=2v_+^2/\lambda_e$ the stability conditions are only narrowly fulfilled, but the real part of $s_1$ stays negative for all the values of $k$. Lastly, when the cell running speeds are not isotropic, the imaginary part of $s_1$ can be non-zero, meaning that there can be stable periodicity in the system. 
While the real part of the other two solutions $s_2$ and $s_3$ is always negative, their imaginary part is non-zero for large values of $k$. More specifically, for large $k$ and finite $D$ their imaginary part is proportional to $k$, while the real part goes with $-Dk^2$. A finite imaginary part indicates oscillations in the system, although the negative real part means that these oscillations are only transient. Signatures of these oscillations can also be seen in our numerical solutions (see the next Section).

\subsection{Repulsion among settled cells}

\begin{figure}[!htbp]
 \begin{center}
 \includegraphics[width=.4\textwidth]{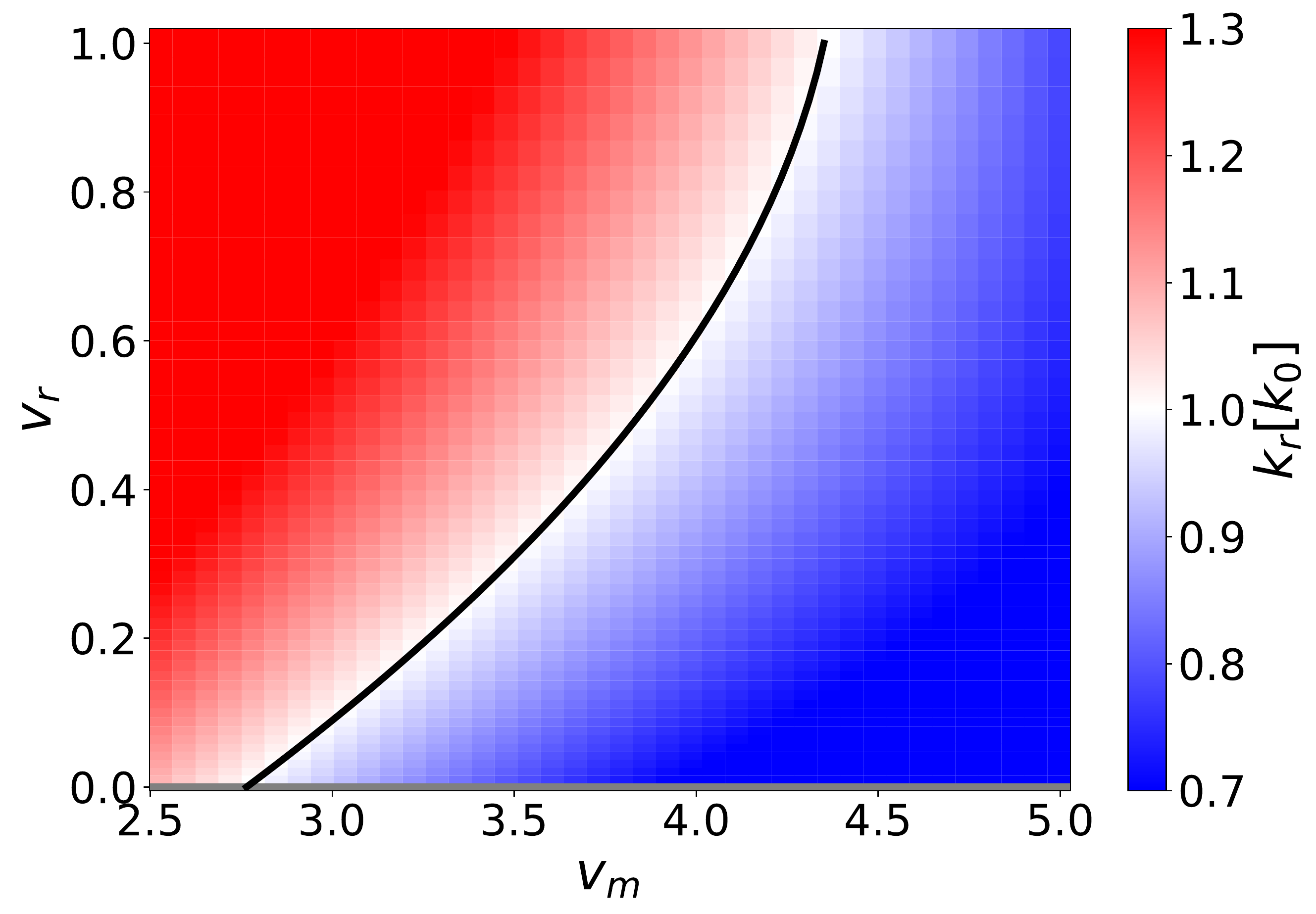}
 \caption{Largest root of $s_1(k)$, $k_r$, as function of $v_r$ and $v_m$. As parameters we chose $\lambda_s =\lambda_d =\mu = 0.1\lambda_e$, $\kappa = 0.2\lambda_e^{-1}$, $\kappa_0 = 0.05\lambda_e^{-1}$ and $D=0.001L^2\lambda_e$. In blue we see the parameters for which the system is not large enough to enable instabilities, while
 in black we have the second order polynomial that fits the $k_r=k_0$ curve.}
\label{Fig8}
\end{center}
\end{figure}

We now include a self-repulsive potential for the cells that do not move, given by $\kappa_0 > 0$ in the matrix $\mathcal{U}$ in Eq. \ref{gee}.  
This repulsion models the need for settled bacteria to not overcrowd any particular region and deplete its resources while reproducing. What is particularly interesting about having 
both an attractive and a repulsive part in the potential is that the interplay of these two opposing effects can lead to structures forming in the system, as we will show now.
If we repeat the analysis described in the last subsection including $\kappa_0>0$, we find that the limits of $s_1(k)$ are
\begin{eqnarray}
s_1(0)&=&\mathcal{E}_{3}\nonumber\\
s_1(k\rightarrow \infty)&=& -\kappa_0\hat{\rho_0} k^2+\mathcal{O}(k).
\end{eqnarray}
The main difference with Eqs.(\ref{19}),(\ref{20}) is that $s_1$ will always be negative for a sufficiently large value of $k$. This means that if we choose parameters for which $s_1$ can be positive,
its largest root $k_r$ will indicate the smallest allowed instability of the system, with size $l=2\pi/k_r$. We consequently expect instabilities to form for systems of size $L$ larger than $l$. As an example of this we numerically calculated the values of $k_r$ for different values of the running speeds $v_+$ and $v_-$, quantifying their occurrence using two non-dimensional parameters, the maximum speed $v_m$ and the reduced difference speed $v_r$ defined by
\begin{equation}
\label{24}
v_m\equiv\frac{\text{max}(v_+,v_-)}{\sqrt{D\lambda_e} }\qquad v_r\equiv \frac{v_+-v_-}{v_++v_-}.
\end{equation}
We chose specifically to vary the running speeds as they can easily tune the asymmetry of the system, leading to interesting instabilities. In Figure \ref{Fig8} we can see $k_r$ as a function of $v_r$ and $v_m$, written in units of $k_0=2\pi/L$. We expect the system to develop instabilities for values of $k_r > k_0$, so we fitted the separatrix $k_r = k_0$ to a second-order polynomial, $v_m^f(v_r)$:
\begin{equation}
\label{25}
v_m^f=2.76\pm0.01+(2.73\pm 0.03)v_r-(1.14\pm 0.04)v_r^2.
\end{equation}
This particular fit was determined using the linear growth model for the parameter values indicated in the caption to Figure \ref{Fig8}.
\\

\begin{figure*}[!htbp]
 \begin{center}
 \includegraphics[width=.9\textwidth]{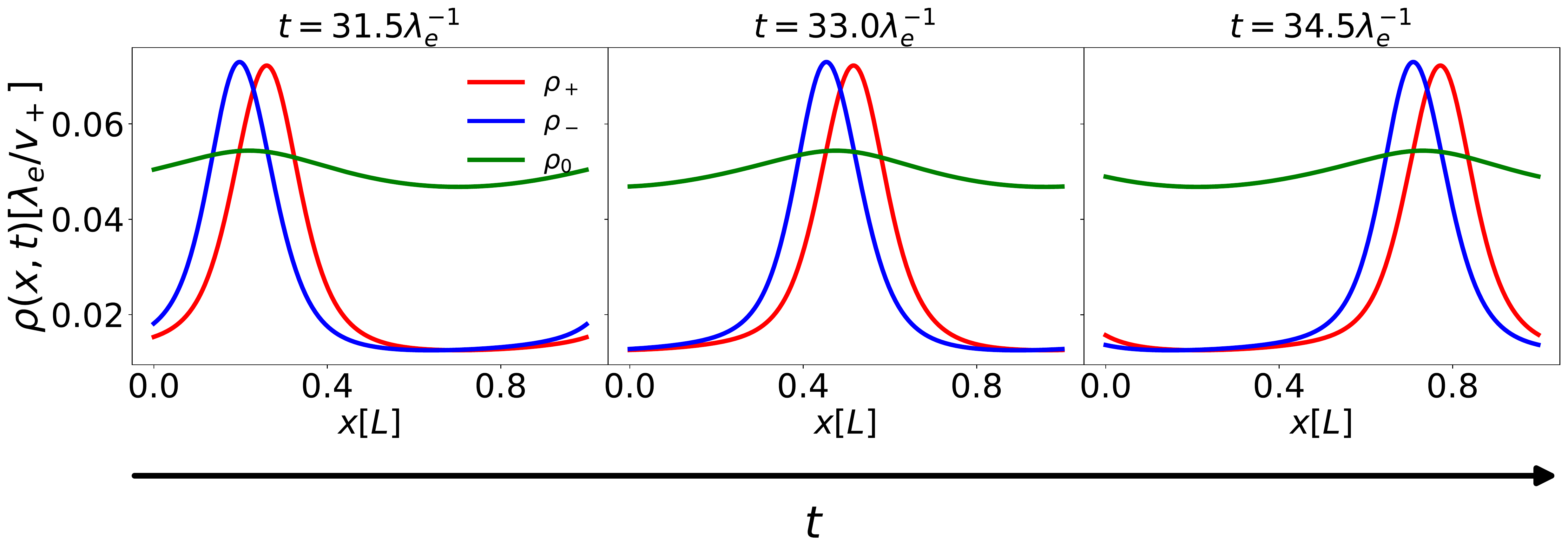}
 \caption{Density of left $\rho_-$, right $\rho_+$ and sedentary $\rho_0$  cells as functions of space at different times (increasing from (a) to (c)). We set here $\lambda_s =\lambda_d =\mu =.1\lambda_e$, $\kappa =.2\lambda_e^{-1}$, $\kappa_0 =.05\lambda_e^{-1}$, $v_+ =2v_- =.1L\lambda_e$ and $D=0.001L^2\lambda_e$.}
\label{Fig9}
\end{center}
\end{figure*}

In order to study the emergence of such instabilities in detail, we further implemented a numerical solver for both Eqs.(\ref{gee}) and (\ref{gee3}), using an explicit fourth-order Runge-Kutta algorithm \cite{press_numerical_2007}  for the time integration and a finite difference scheme in space. We performed calculations both with the linear and the nonlinear grow\-th model, setting respectively 
$\lambda_s=\mu$ and $\lambda_s\geq \mu$.
We use a finite box of length $L$ with periodic boundary conditions. Setting the time step to $\Delta t=10^{-4}\lambda_e^{-1}$ we calculated $\sim 10^6$ steps
to ensure that the system settles into a steady state.
Our calculations are initialized using the steady-state solutions of the linear system (Eqs.~\eqref{7}), to which we add small fluctuations given by Gaussian noise.
We find that our system develops wave-like structures, which are static for $v_+=v_-$ and become traveling waves when $v_+\neq v_-$ - see  Figure~\ref{Fig9} for the linear growth case and  Figure~\ref{Fig11} for the nonlinear case. Testing different initial conditions, e.g. choosing $\rho_0(x)$ as a narrow Gaussian peak that approximates an initially settled single cell,
we also observed that these wave-like structures always form, even if the specific shape of the wave can be affected.
In our analysis we preferred to use the steady-state solution of Eqs.(\ref{7}) as initial condition, as it makes comparison with the theoretical results of Figure \ref{Fig8} more straightforward. Intuitively, the attractive term $\kappa$ leads to the formation of peaks, induced by the instability in Eq.(\ref{stability}). These peaks are then stabilized by the repulsive term $\kappa_0$. The asymmetry of the running speeds makes the peaks move.

\begin{figure}[!htbp]
 \begin{center}
 \includegraphics[width=.4\textwidth]{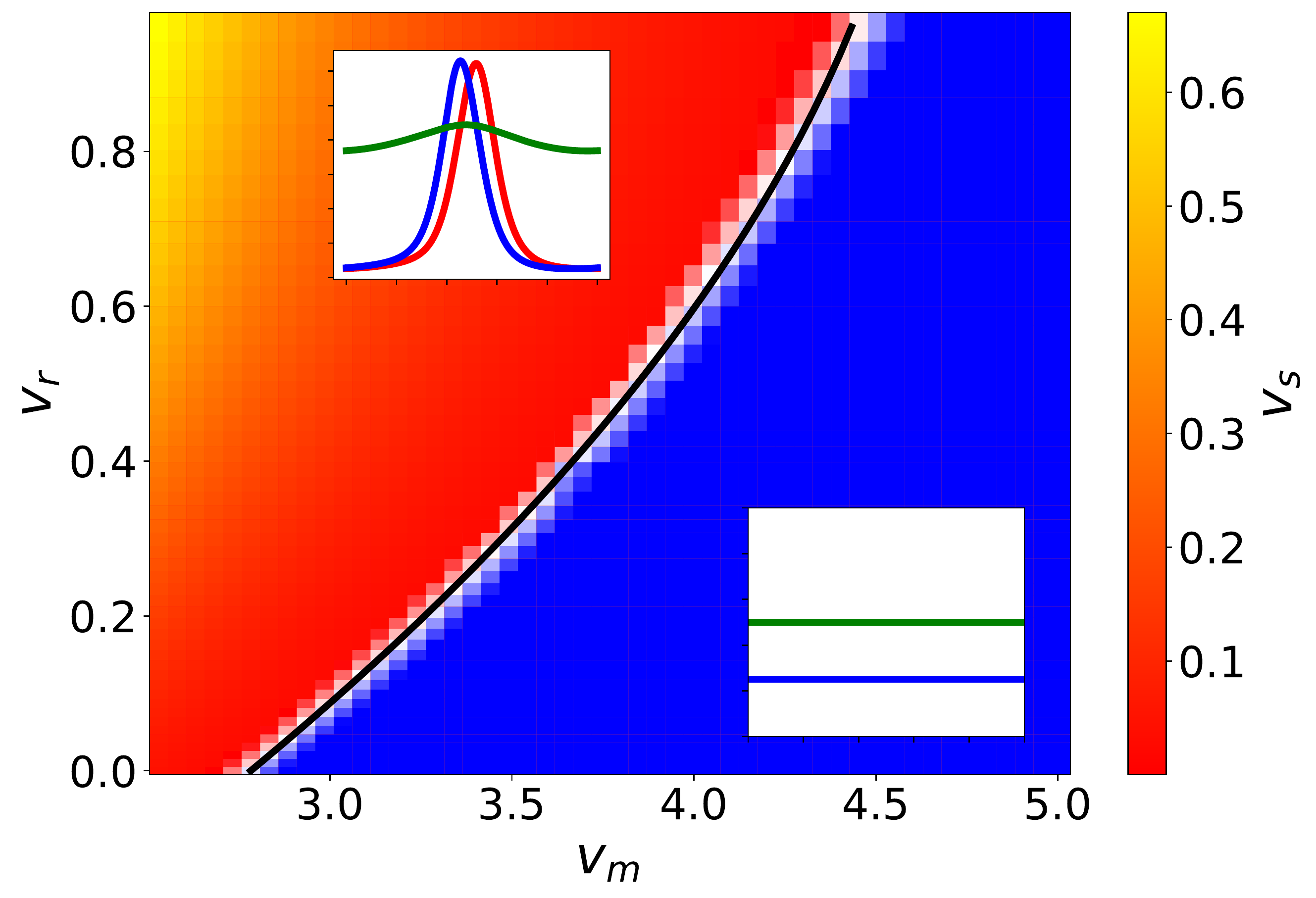}
 \caption{State diagram of the system as a function of $v_r$ and $v_m$. As parameters we chose $\lambda_s =\lambda_d =\mu = 0.1\lambda_e$, $\kappa = 0.2\lambda_e^{-1}$, $\kappa_0 = 0.05\lambda_e^{-1}$ and $D=0.001L^2\lambda_e$. In blue we see the parameters for which the system is stably constant, while in red to yellow we see the parameters for which the system 
 generates traveling wave structures. Examples of both long-time behaviors are shown in their respective area. The gradient shows the stationary velocity of the waves $v_s$, while
 in black we have the second order polynomial that fits the transition curve $v_m^f$.}
\label{Fig10}
\end{center}
\end{figure}

\begin{figure*}[!htbp]
 \begin{center}
 \includegraphics[width=.9\textwidth]{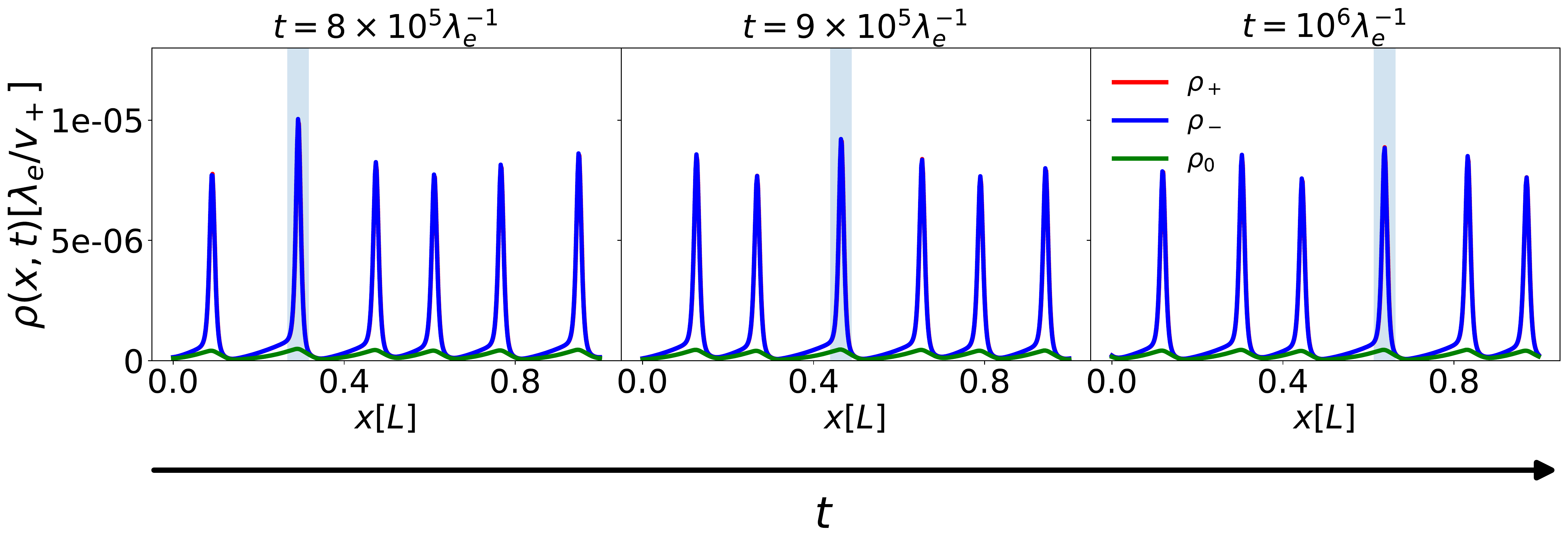}
 \caption{Nonlinear growth model, density of left $\rho_-$, right $\rho_+$ and sedentary $\rho_0$  cells as functions of space at different times (increasing from (a) to (c)). Because of the large value of $\lambda_e$ compared to the other rate parameters, the right-moving and left-moving populations have almost the same shape, making the red line disappear under the blue line. The shaded areas indicate the largest peak, and how it moves in time towards the right. We chose as parameters the values typical of {\it CC} shown in Table \ref{Table1}.
For the interaction potentials, for which no experimental estimates can be made at present, we chose $\kappa =\kappa_0=10\lambda_e^{-1}$, while for the carrying capacity of the system we set 
$\boldsymbol\rho_c(x,t)=(0.04,0.04,0.04)\lambda_e/v_+$.}
\label{Fig11}
\end{center}
\end{figure*}

Migrating bands of bacteria have indeed been observed experimentally  \cite{adler_chemotaxis_1966,adler_effect_1966,adler_chemoreceptors_1969,berleman_rippling_2006,stricker_hybrid_2020,liu_viscoelastic_2021} 
and have also been modeled theoretically \cite{cremer_chemotaxis_2019,narla_traveling-wave_2021,caprini_collective_2021}, always considering only one species of cells.
A particularly surprising feature of our model is that in this final stationary state all three distributions evolve in the same direction at the same speed, independently of the intrinsic running speed of the 
cells.
We replicated the diagram of Figure \ref{Fig8} with numerical integration of the model equation, and the  
resulting non-equilibrium state diagram is shown in Figure \ref{Fig10}. We find a clear transition from a stable system (shown in blue), where all species are constant in space, to the appearance of wave-like structures (shown in red to yellow). The gradient visualizes the change in stationary speed of the waves $v_s$, defined as the speed of the waves in the stationary state divided by $\sqrt{D\lambda_e}$, and is hence non-dimensional. This quantity is almost vanishing near the transition, and grows the further away we move from it. The formation of these waves is typical of systems with a large difference between $v_+$ and $v_-$ or rather small absolute speeds. We fitted the separatrix to a second order polynomial $v_m^f(v_r)$ and obtained
\begin{equation}
v_m^f=2.78\pm0.01+(2.56\pm 0.03)v_r-(0.88\pm 0.03)v_r^2.
\end{equation}
We find that our numerical calculations and theory are in very good qualitative agreement.

\subsection{Application to {\it Caulobacter crescentus}}

Table \ref{Table1} gives an idea of the experimentally measured values for {\it CC} which have been extracted from recent papers on its swimming behaviour
\cite{li_low_2006,lin_single-gene_2010,liu_helical_2014,lele_flagellar_2016}. It is noteworthy to comment on the running speeds $v_+, v_-$. While the torque generated by the flagellar motor differs significantly during forward and backward motion, the resulting velocities are not dramatically different (and, in fact, experimentally hard to measure) \cite{lele_flagellar_2016}. 
\begin{table} 
\begin{center}
\begin{tabular} {|l|c|c|}
\hline
\mbox{run-and-tumbling rate}; $s^{-1}$ & $\lambda_e$ & $10^{-1}$ \\
\hline
\mbox{settling rate}; $s^{-1}$ & $\lambda_s$ & $10^{-5}$ \\
\hline
\mbox{doubling rate}; $s^{-1}$ & $\lambda_d$ & $10^{-4}$ \\
\hline
\mbox{decay rate}; $s^{-1}$ & $\mu$ & $10^{-6}$ \\
\hline
\mbox{running speed right}; $m/s$ & $v_+$ & $4 \cdot 10^{-5}$  \\
\hline
\mbox{running speed left}; $m/s$ & $v_- $ & $3.5 \cdot 10^{-5}$ \\
\hline 
\mbox{diffusion coefficient}; $m^2/s$ & $D$ & $2 \cdot 10^{-9}$ \\
\hline
\end{tabular}
\end{center}
\caption{Values of the parameters for {\it Caulobacter crescentus} taken from \cite{li_low_2006,lin_single-gene_2010,liu_helical_2014,lele_flagellar_2016}.}
\label{Table1}
\end{table}
We have performed calculations with the parameters of Table \ref{Table1} for different values of $\kappa$ and $\kappa_0$ which are undetermined from experiments. 
Since for {\it Caulobacter} $\mu < \lambda_s$, we have included the saturating nonlinearity for the growth in the model. The results show that the 
waves still form provided the ratio $\kappa/\kappa_0$ is large enough (Figure \ref{Fig11}).

\section{Conclusions and outlook}
\label{Sec:conclusions}

In this work we proposed and studied a 1D 3-state model motivated by the cell cycle progression of the bacterium {\it Caulobacter crescentus}, including both its run and tumble 
motion and its reproductive behavior. We first  analyzed the free cell space-independent case and calculate the parameter regimes for which the number of cells grows or 
declines. Adding the spatial dependence we subsequently determined dynamical quantities of the system such as the mean displacement, the mean-squared displacement and 
the intermediate scattering function. 
We found a surprising super-ballistic behavior of the MSD at short times with a $t^3$ scaling which stems from the interplay of cells doubling and cells starting to swim. 

Subsequently, we included attractive and repulsive interactions between cells into our model, representing their tendency to swim towards regions in which cells are settled
and to avoid overcrowding. We determined the stability conditions and, using numerical methods, we studied the fully nonlinear system in which we 
identify traveling waves of cells. Their occurrence is quantified in a non-equilibrium state diagram.

Our model lends itself to further extensions in several ways. E.g., one could account for complex nutrient landscapes and for a more detailed description of the cell cycle, which is well-studied from various
aspects \cite{biondi_cell_2022}; another possible system for application are \textit{Chlamydomonas reinhartii} cells \cite{harris_chlamydomonas_2009}.  The cell cycle can be included in cell-resolved simulations such as performed recently in \cite{you_geometry_2018,schwarzendahl_notitle_2022}. Another direction could be a two-dimensional field description that includes the nematic ordering of cells such as in \cite{dellarciprete_growing_2018}. In a higher-dimensional model it would also be interesting to see what the effect of different swimming strategies such as run and tumble, run-reverse or run-reverse-flick~\cite{grognot_more_2021} is. Finally, an exploration of the fully nonlinear model - nonlinear diffusive interactions as well as nonlinear growth - including a full higher-dimensional 
tumbling behaviour for a multi-species system would be an interesting problem in the context of biofilm growth.



\section*{Appendix}

\subsection*{Relation between intermediate scattering function and momenta of the density in 1D}

We show here the calculation that justifies Eq.(\ref{11}) in one dimension in the case where the initial conditions for the cell density are $\boldsymbol\rho(x,t=0) = (0,N(0)\delta(x),0)$. First, we write the definition for the moments $\langle (x(t)-x_0)^n\rangle=\langle x^n(t)\rangle$:
\begin{equation}
\langle x^n(t)\rangle=\int_{-\infty}^\infty \text{d}x\,  x^n P(x,t),
\end{equation}
where $P(x,t)$ is the probability density of the position. We then apply a Fourier transform and its inverse in the integral
\begin{equation}
\langle x^n(t)\rangle=\frac{1}{2\pi}\int_{-\infty}^\infty \text{d}x \int_{-\infty}^\infty \text{d}k\,\text{e}^{ikx}\left( \text{i}^n \frac{\partial ^n \tilde{P}(k,t) }{\partial k^n} \right)  ,
\end{equation}
where $\tilde{P}(k,t)$ is the Fourier Transform of $P(x,t)$. Finally, we exchange the order of integration to get
\begin{eqnarray}
\langle x^n(t)\rangle &=&\frac{1}{2\pi}\int_{-\infty}^\infty \text{d}k\, 2\pi\delta(k)\left( \text{i}^n \frac{\partial ^n \tilde{P}(k,t) }{\partial k^n} \right)\nonumber\\ &=&\left.\text{i}^n \frac{\partial ^n \tilde{P}(k,t)}{\partial k^n}\right|_{k=0}.
\end{eqnarray}
Knowing that for the initial conditions that we have chosen $\tilde{\boldsymbol\rho}(-k,0)=(0,N(0),0)$, we have
\begin{equation}
\mathcal{F}(k,t)\equiv \tilde{P}(k,t)\tilde{P}(-k,0)N(t) = \tilde{P}(k,t)N(t),
\end{equation}
and hence
\begin{equation}
\langle x^n(t)\rangle=\left.\text{i}^n \frac{\partial ^n \tilde{P}(k,t)}{\partial k^n}\right|_{k=0}=\left.\frac{\text{i}^n}{N(t)} \frac{\partial ^n\mathcal{F}(k,t)}{\partial k^n}\right|_{k=0}.
\end{equation}

\section*{Acknowledgements}
DB is supported by the EU MSCA-ITN ActiveMatter, (proposal No. 812780). RB is grateful to HL for the invitation to a stay at the Heinrich-Heine-University in 
D\"usseldorf where this work was performed. HL was supported by the DFG project LO 418/25-1 of the SPP 2265. 

\section*{Author contribution statement}
HL and RB directed the project. DB performed analytic calculations
and
numerical simulations. All authors discussed the results
and
wrote the manuscript.

\section*{Data Availability Statement}
The datasets generated during and/or analysed during the current study are available from the corresponding author on reasonable request.

\bibliography{Caulobacter}

\end{document}